\documentclass[12pt]{iopart}

\usepackage{graphicx}
\usepackage{iopams}
\usepackage{color}

\begin{document}

\title{`Quantum Cheshire Cat' as Simple Quantum Interference}

\author{Raul Corr\^ea, Marcelo Fran\c ca Santos, C H Monken and Pablo L Saldanha} 
\address{Departamento de F\'isica, Universidade Federal de Minas Gerais, Caixa Postal 702, 30161-970, Belo Horizonte, MG, Brazil}

\ead{raulcs@fisica.ufmg.br}

\begin{abstract}
In a recent work, Aharonov \textit{et al.} suggested that a photon could be separated from its polarization in an experiment involving pre- and post-selection [New J. Phys \textbf{15}, 113015 (2013)]. They named the effect `quantum Cheshire Cat', in a reference to the cat that is separated from its grin in the novel \textit{Alice's Adventures in Wonderland}. Following these ideas,  Denkmayr \textit{et al.} performed a neutron interferometric experiment and interpreted the results suggesting that neutrons were separated from their spin [Nat. Commun. \textbf{5}, 4492 (2014)]. Here we show that these results can be interpreted as simple quantum interference, with no separation between the quantum particle and its internal degree of freedom. We thus hope to clarify the phenomenon with this work, by removing these apparent paradoxes. 
\end{abstract}


\maketitle

\section{Introduction}
The concept of a quantum weak value, introduced in 1988 by Aharonov, Albert, and Vaidman \cite{aharonov88}, has allowed the development of novel and important experimental techniques to study quantum systems~\cite{dressel14}. The amplification of small signals beyond technical noise \cite{hosten08,dixon09}, the direct determination of quantum states \cite{lundeen11} and geometric phases~\cite{sjoqvist06}, and the characterization of the nonclassical behavior of quantum systems \cite{palacios10} are a few examples of its usefulness (check  \cite{dressel14} for a more extensive list of applications). Together with the new interferometric concept, the original paper and some follow-ups also suggested that weak measurements could lead to a new interpretation of quantum phenomena. These somewhat controversial ideas, however, were the target of lots of discussions as can be seen in the comments following the 1988 original paper~\cite{leggett89,peres89,aharonov89,duck89}.

Recently, a new set of proposals and experiments has revived some of this controversy by suggesting even more radical ways of reinterpreting quantum mechanics {\cite{aharonov13,bancal14,denkmayr14,lorenzo12}}. For instance, in a recent theoretical work, Aharonov \textit{et al.} argued that a particular weak measurement setup for photons allowed one to state that ``in the curious way of quantum mechanics, photon polarization may exist where there is no photon at all'' \cite{aharonov13}. The idea was reinforced both in a \textit{news} \& \textit{views} article of Nature Physics~\cite{bancal14} and in an experiment performed with neutrons~\cite{denkmayr14}. The first discusses the results of \cite{aharonov13} concluding that ``polarization could be effectively isolated from the photons carrying it'', while the second implements an equivalent interferometer to the one proposed in \cite{aharonov13} and argues that ``The experimental results suggest that the system behaves as if the neutrons go through one beam path, while their magnetic moment travels along the other'' \cite{denkmayr14}. The phenomenon was nicknamed ``quantum Cheshire Cat'', in a reference to the cat that is separated from its grin in the novel \textit{Alice's Adventures in Wonderland}, by Lewis Carroll.

{According to \cite{aharonov88}, in order to obtain a weak value one must interact the system to be measured with a continuous degree of freedom described by a wavefunction, which will act as a quantum pointer. Before the interaction, the system is pre-selected in a known state $|\Phi_{\mathrm{pre}}\rangle$ and, after interacting with the pointer, a post-selection is made on a specific state of the measured system $|\Phi_{\mathrm{post}}\rangle$, finally preparing the pointer in a state that depends on both $|\Phi_{\mathrm{pre}}\rangle$ and $|\Phi_{\mathrm{post}}\rangle$. Given the weakness of the interaction, the wavefunction of the pointer is displaced by a small amount to which the weak value is proportional.} 

Most of the controversy in trying to extract a new interpretation out of weak measurements lies in the attempt to attach physical reality to the otherwise mathematically defined weak values. {In \cite{aharonov13}, the continuous degrees of freedom acting as quantum meters are not considered in the construction of the argument for the new interpretation.} In the present work we show that by taking these degrees of freedom into account, both the theoretical predictions and experimental results that motivated the somewhat unusual ``Cheshire Cat'' interpretation can be explained as simple quantum interference where no detachment between the photon and its polarization or between the neutron and its magnetic moment is actually required. We begin by quickly revisiting the proposal of \cite{aharonov13}, then we proceed to explain it with standard quantum mechanics interference and we finally show how a similar approach also explains the experiment described in~\cite{denkmayr14}. 

\section{The Aharonov \textit{et al.} proposal}
\begin{figure}
\centering
  \includegraphics[width=10cm]{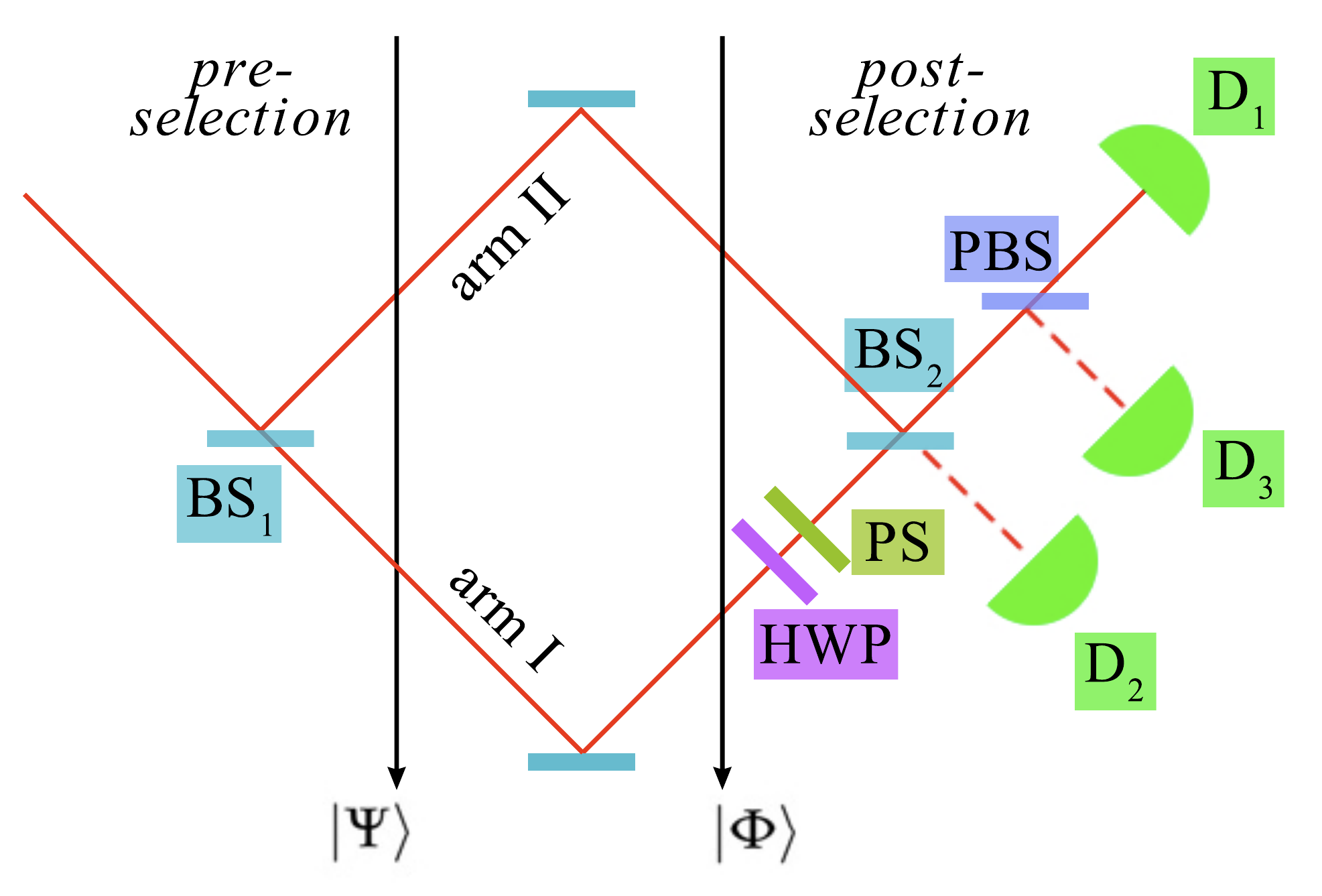}\\
  \caption{Interferometer setup from \cite{aharonov13}. The photon reaches a beam splitter (BS$_1$) with polarization $H$, right after which we have the state $|\Psi\rangle$ from Eq. (\ref{psi}), called \textit{pre-selection state}. After reflection from mirrors on both arms $I$ and $II$, we want to make the \textit{post-selection} of the state $|\Phi\rangle$ from (\ref{phi}) on detector D$_1$. A half-wave plate (HWP) interchanges $H$ and $V$ polarizations on arm $I$ and a phase shifter (PS) adds a specific phase right after it, so that, after a second beam splitter (BS$_2$) and a polarizing beam splitter (PBS) that transmits only horizontal polarization, we assure that D$_1$ will always click for $|\Phi\rangle$ and never click for any state orthogonal to it.}
\label{fig:interf}
\end{figure}

In \cite{aharonov13}, the authors base their proposal in the interferometer shown in figure \ref{fig:interf} which is designed to prepare the photon in state $|\Psi\rangle$ and post-select it in state $|\Phi\rangle$, both specified below:
\begin{equation}
\label{psi}
|\Psi\rangle=\frac{1}{\sqrt{2}}\Big(|I\rangle+\rmi|II\rangle\Big)|H\rangle,
\end{equation}
\begin{equation}
\label{phi}
|\Phi\rangle=\frac{1}{\sqrt{2}}\Big(|I\rangle|V\rangle+|II\rangle|H\rangle\Big),
\end{equation}
where $|I\rangle$ and $|II\rangle$ correspond to the photon being in each arm of the interferometer shown in figure \ref{fig:interf}, and $|H\rangle$ and $|V\rangle$ represent its horizontal and vertical polarization components. The left and right circular polarization states can be written as ${|\pm\rangle=(|H\rangle\pm \rmi|V\rangle)/\sqrt{2}}$ and these are the eigenstates of the ``spin'' angular momentum operator for the photon.

In order to do the desired pre- and post-selection, beam splitters (BS$_1$ and BS$_2$), a half-wave plate (HWP), a phase shifter (PS) and a polarizing beam splitter (PBS) are placed and adjusted in a way that a photon on the state $|\Phi\rangle$ right before HWP will always click at detector D$_1$; if it is in any state orthogonal to $|\Phi\rangle$, it necessarily clicks elsewhere. The adjustments of the devices are described in \cite{aharonov13}.

After being prepared in the state $|\Psi\rangle$, the photon interacts with devices positioned in both interferometer paths. These devices act as probes and can perform either projective measurements (also called ``strong'') or weak measurements. Although in \cite{aharonov13} the authors considered them as quantum devices, the description of their quantum state is not included in the paper. We will explicitly represent their quantum states here, though, as this is the essential piece for describing the phenomenon as quantum interference. In arm $I$ there is a polarization detector whose quantum state $|P_1\rangle$ changes to $|P_1^+\rangle\ (|P_1^-\rangle)$ if a photon of polarization $|+\rangle\ (|-\rangle)$ propagates through that arm, or stays at $|P_1\rangle$ if a photon propagates through arm $II$. Meanwhile, in arm $II$ there is a quantum device to measure the presence of the photon, i.e. a device whose quantum state $|P_2\rangle$ changes into $|P_2^+\rangle$ if a photon propagates through that arm, or stays at $|P_2\rangle$ if a photon propagates through arm $I$. In the proposal of \cite{aharonov13} the probe of arm $I$ is a combination of birrefringent materials that produces a positive horizontal displacement for a photon with polarization $|+\rangle$ and a negative horizontal displacement for a photon with polarization $|-\rangle$. The pointer state $|P_1\rangle$ of the device is thus associated to the center of the beam in the horizontal direction. The probe in arm $II$ is a glass sheet that displaces the photon beam up. The pointer state $|P_2\rangle$ of the device is thus associated to the center of the beam in the vertical direction. If the beam displacements produced by the measurement devices are greater than the beam diameter, then we have a projective measurement where we can associate the vertical displacement with the photon path and the horizontal displacement with the polarization of the photon propagating through arm $I$. If the beam displacements are much smaller than the beam diameters, then we are in the weak measurements regime. In this case we have, for instance, $\langle P_1|P_1^+\rangle\approx1-\epsilon$ with $|\epsilon|\ll1$, as opposite to $\langle P_1|P_1^+\rangle=0$ for a projective measurement. 

After the interaction with the measurement devices in each path, the composite state of the photon and the pointers is
\begin{eqnarray}\nonumber
	|\Psi'\rangle=&&C\Bigg[\frac{1}{2}|I\rangle|+\rangle|P_1^+\rangle|P_2\rangle+\frac{1}{2}|I\rangle|-\rangle|P_1^-\rangle|P_2\rangle+\\
	\label{interaction}
	&&+\frac{\rmi}{\sqrt{2}}|II\rangle|H\rangle|P_1\rangle|P_2^+\rangle\Bigg].
\end{eqnarray}
{Note that the changes in the state of the pointer are only defined by the shapes of the wavefunctions of the probes and by their displacements and, therefore, are unaffected by global phases and normalization factors. 
For that reason we have just added a factor $C$ which accounts for these global factors (normalization and phase) but does not influence by any means the overall analysis.} 

To make a post-selection of the results only for photons that leave the PBS in direction to D$_1$ means to project the above state (\ref{interaction}) on $|\Phi\rangle$ from (\ref{phi}). We end up with the following quantum state for the pointers
\begin{equation}
\label{projection}
\langle\Phi|\Psi'\rangle=\frac{\rmi C}{4}\bigg[2|P_1\rangle|P_2^+\rangle+|P_1^+\rangle|P_2\rangle-|P_1^-\rangle|P_2\rangle\bigg].
\end{equation}
Note that in the state (\ref{projection}) there is entanglement between the pointers, {meaning that there are quantum correlations between the measurement devices placed in arms $I$ and $II$, even though they have never interacted directly with each other \cite{lorenzo14}.} If the devices make projective measurements, there are three possibilities for each photon{: either the photon has an up displacement greater than the beam diameter or the photon has a positive (negative) horizontal displacement greater than the beam diameter. In these cases, if one assumes a real trajectory for the photon, the first situation is compatible with propagation through arm $II$, and the second (third) is compatible with propagation through arm $I$ and polarization $|+\rangle$($|-\rangle$).} 
{On the other hand, when the measurement devices interact weakly with the photon, producing vertical and horizontal displacements much smaller than the beam diameter, such assumptions about past trajectories of the photon simply cannot hold true anymore due to the quantum interference between the different possible paths. In fact, it is well known that  attributing physical reality to the past of quantum particles inside interferometers leads to paradoxes \cite{gibbins}. However, that is exactly what the authors of \cite{aharonov13} do when they extend the interpretation used in the projective measurements to the weak interaction case.} They consider that if the average vertical displacement of a set of photons is the same {as the displacement} of the wavefunction of one photon eventually propagating through arm $II$, then this indicates that the photons had propagated through this arm. In the same way, they also assume that if the average horizontal displacement of a set of photons is the same of the wavefunction of one photon eventually propagating through arm $I$ with polarization $|+\rangle$, then there is a $|+\rangle$ polarization in arm $I$. This is the origin of the paradox when concluding that ``the photon is in the left arm (...) while the angular momentum is in the right arm'' \cite{aharonov13}. As we show in the following, there is no such paradox if we describe the phenomenon as simple quantum interference.

Let us now explicitly describe the pointers as suggested in \cite{aharonov13}, which will be the transversal beam profiles of the photon in the fashion of a Bialynicki-Birula--Sipe photon wave function \cite{birula94,birula96,sipe95,saldanha11} in the paraxial regime. In this regime, the beam propagation is highly directional, such that its state can be written as the product of a transversal function in the $xy$ plane (with $z$-dependent parameters) and a function that describes its evolution while propagating in the $z$ direction \cite{mandel}. The beam polarization can also be treated as an independent parameter in this regime. We may consider that the transversal properties of the beam do not vary much with the propagation through the interferometer, such that the transversal wave function can be written as a function only of the transversal components $x$ and $y$. In the proposal of \cite{aharonov13}, $|P_1\rangle$ corresponds to a wave function $f(x)$ and $|P_2\rangle$ corresponds to $g(y)$, such that $|P_1^+\rangle$, $|P_1^-\rangle$ and $|P_2^+\rangle$ correspond to the same functions displaced by $\delta_x$, $-\delta_x$ and $\delta_y$, respectively. Hence, by defining $F(x,y)\equiv f(x)g(y)$, the state (\ref{projection}) is described by the transversal wave function:
\begin{equation}
\label{wavefunction}
F_1(x,y)=C_{1}[2F(x,y-\delta_y)+F(x-\delta_x,y)-F(x+\delta_x,y)],
\end{equation}
{where $C_{1}$ again is a normalization factor that depends on how much the components of (\ref{wavefunction}) overlap with each other and on the shape of $F(x,y)$.} If the detector D$_1$ is a screen sensitive to the photon position, after many runs we can build the distribution $|F_1(x,y)|^{2}$ on it. The pattern showed will depend on how orthogonal these three displaced functions above are, and how their overlap creates interference. In a \textit{strong} (projective) measurement regime, the interaction displaces each term above the beam radius, such that their overlap is negligible and there is no interference. Each part of the function is then identifiable on the screen, offering trustworthy information about the measured quantities. {This behaviour is clear when we realize that, without significant overlap of the pointer states, that is for instance $\langle P_1|P_1^+\rangle\rightarrow 0$, the probability density is approximately}
\begin{equation}
\label{prob-dens}
|F_1(x,y)|^2\approx|C_{1}|^2[|2F(x,y-\delta_y)|^2+|F(x-\delta_x,y)|^2+|F(x+\delta_x,y)|^2].
\end{equation}
The \textit{weak measurement} will happen in the case that the displaced functions are nearly completely overlapping. They hence interfere {and $|F_1(x,y)|^2$ cannot be approximated by (\ref{prob-dens})} -- it may even look like the same function displaced by some other amount. But, as we discussed before, the reason of the paradoxes in \cite{aharonov13,denkmayr14} is to consider that the resultant displacement may be read as a measurement in the ordinary sense.

Using the approximations $F(x\pm\delta_x,y)\approx F(x,y)\pm\delta_x\frac{\partial F(x,y)}{\partial x}$ and $F(x,y-\delta_y)\approx F(x,y)-\delta_y\frac{\partial F(x,y)}{\partial y}$ in (\ref{wavefunction}), we end up with
\begin{equation}
\label{approx}
F_1(x,y) \approx 2C_{1}F(x-\delta_x,y-\delta_y).
\end{equation}

According to the arguments of \cite{aharonov13,denkmayr14}, this result is compatible with the situation where the photon is measured in the left arm of the interferometer (the beam was displaced up by $\delta_y$) and, at the same time, there is positive angular momentum on the right arm of the interferometer (the beam was displaced sideways by $\delta_x$). However, as pointed out in our calculations, the probability of finding the photons at this particular range of positions can be interpreted as simple interference. Note that any quantum continuous variable could have been used as a probe. The photon wavefunction is a particularly effective example because it fulfils both the superposition principle and the approximations discussed above. Also note that the same holds true for the Schr\"odinger wavefunction. Actually, our treatment is suitable to describe even the light intensity distribution on the detector if classical electromagnetic waves are sent through the interferometer.

\begin{figure}
\centering
  \includegraphics[width=15cm]{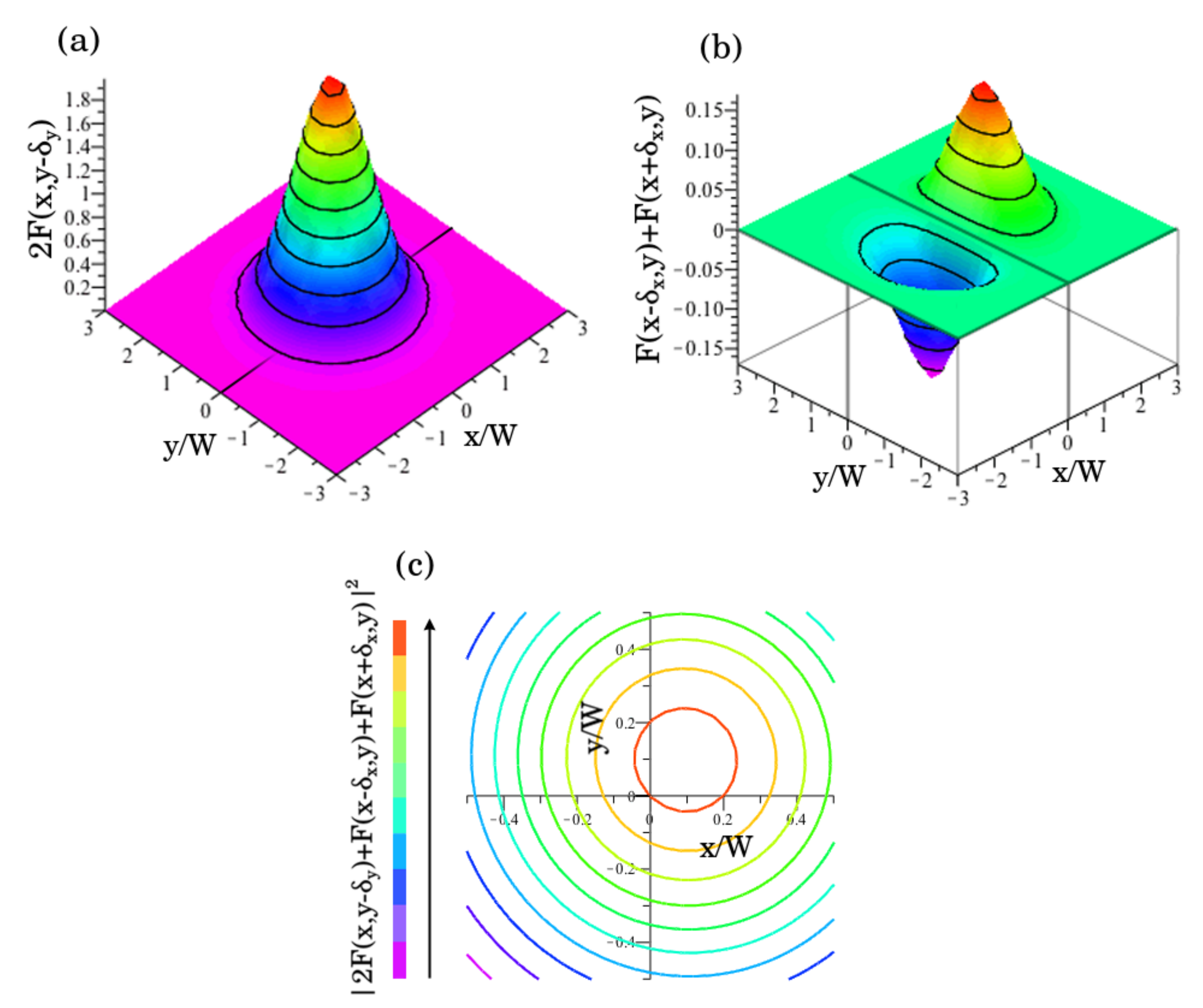}
  \caption{Plots of the non-normalized functions (a) $2F(x,y-\delta_y)$ and (b) $[F(x-\delta_x,y)-F(x+\delta_x,y)]$, for $F(x,y)=e^{-(x^{2}+y^{2})/W^2}$ and $\delta_x=\delta_y=0.1W$. The amplitude should be noted, where in (a) the maximum is around 1.5, and in (b) it's around 0.012. In (c) we plot $|2F(x,y-\delta_y)-F(x-\delta_x,y)-F(x+\delta_x,y)|^{2}$, to show that the displacement is small enough so the pointer indicates the weak values.}
\label{fig:graf}
\end{figure}

To make this issue clearer, in figure \ref{fig:graf}(a) we plot $2F(x,y-\delta_y)=2f(x)f(y-\delta_y)$, which is the component of the photon wavefunction that comes from the arm $II$ in (\ref{wavefunction}), with $f(x)$ being a Gaussian function with width $W$ centred at zero. In figure \ref{fig:graf}(b) we plot $F(x-\delta_x,y)-F(x+\delta_x,y)$, which is the component of the photon wavefunction that comes from arm $I$ in (\ref{wavefunction}).  We can see that for $\delta_x=\delta_y\ll W$ the major contribution to the wavefunction of (\ref{wavefunction}) comes from $2F(x,y-\delta_y)$. But the term $F(x-\delta_x,y)-F(x+\delta_x,y)$ interferes destructively with $2F(x,y-\delta_y)$ for negative $x$ and constructively for positive $x$, resulting in an overall positive displacement in the horizontal direction for the wavefunction even if this term is small, as can be seen in \ref{fig:graf}(c). A similar argument was used in \cite{saldanha14} to present a classical explanation of the experimental results of \cite{danan13}. In \cite{danan13} the authors performed experiments with a nested Mach-Zehnder interferometer and concluded that ``the photons tell us that they have been in the parts of the interferometer through which they could not pass'' \cite{danan13}. But again, this odd conclusion was achieved through the attribution of a physical reality to the weak value of a weak measurement. 

\section{The experiment of Denkmayr \textit{et al.}}
\begin{figure}
\centering
  \includegraphics[width=10cm]{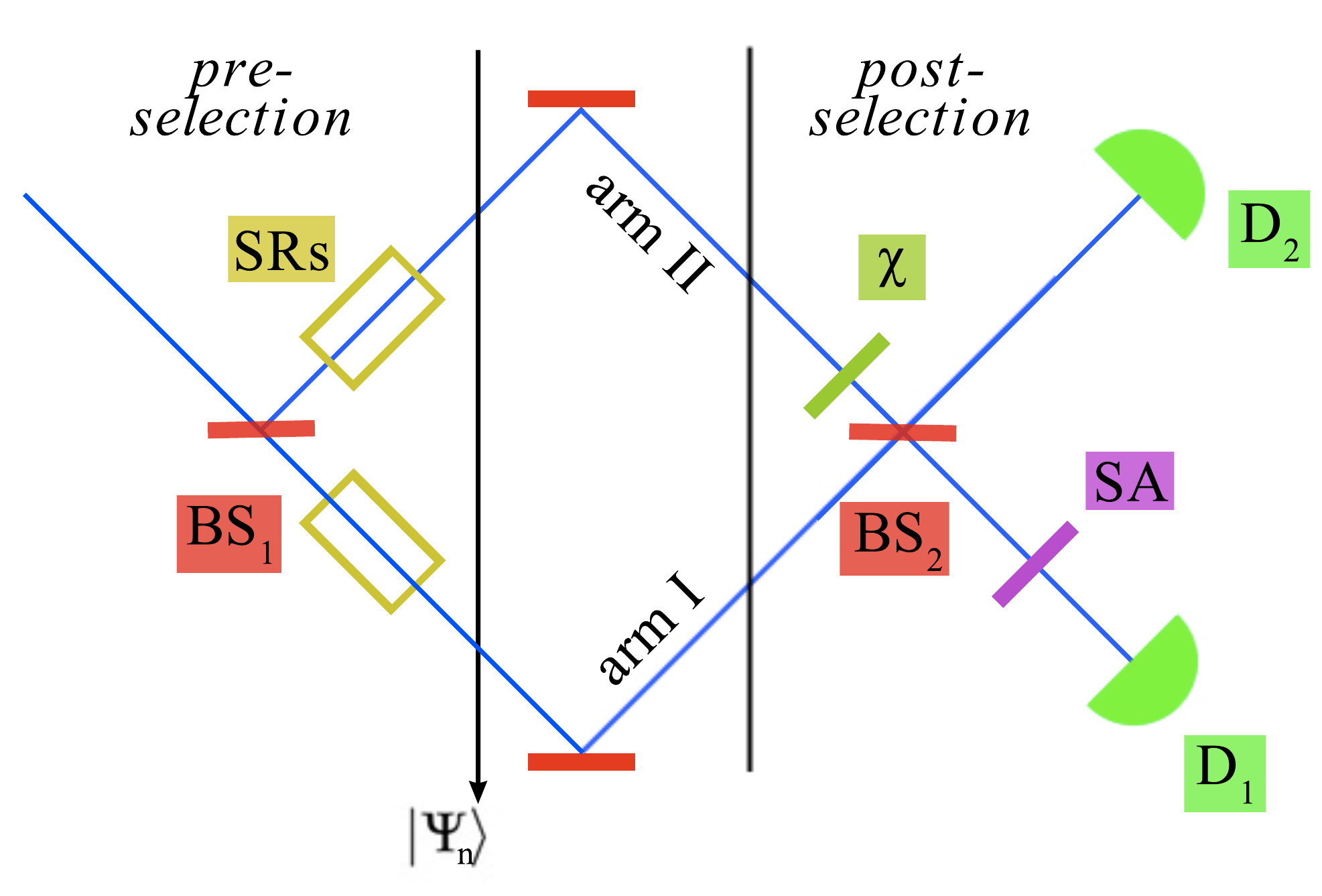}
  \caption{Experimental setup of \cite{denkmayr14}. A neutron beam reaches a beam splitter BS$_1$ and the spin rotators (SR) on each arm $I$ and $II$ are set to leave each neutron on the state $|\Psi_{n}\rangle$ from (\ref{psin}). A phase $\chi$ between the paths may be controlled before the beams are combined at BS$_2$. A spin analyzer (SA) selects only the $|-\rangle$ component of spin to be detected by the neutron detector D$_1$. There is no spin selection for D$_2$. Absorbers may be placed and magnetic fields can be applied to rotate the neutron spin in each path.}
\label{fig:experim}
\end{figure}

The experimental realization by Denkmayr \textit{et al.} \cite{denkmayr14} was similar to the theoretical proposal of \cite{aharonov13} and is depicted in figure \ref{fig:experim}. After going through the first beam splitter and the spin rotators, the neutron is prepared in the state 
\begin{equation}
\label{psin}
|\Psi_{n}\rangle=\frac{1}{\sqrt{2}}\Big(|I\rangle|+\rangle+|II\rangle|-\rangle\Big).
\end{equation}
$|I\rangle$ and $|II\rangle$ represent the neutron states in the corresponding interferometer arm and  $|\pm\rangle$ are the eigenvectors of the neutron $x$-component spin operator. There is a spin analyzer before D$_1$, such that the detection of a neutron in D$_1$ is associated to the projector
\begin{equation}\label{pi1}
	\hat{\Pi}_1=\Big(|I\rangle+\rme^{\rmi\chi}|II\rangle\Big)\Big(\langle I| + \rme^{-\rmi\chi}\langle II|\Big)\otimes|-\rangle\langle-|,
\end{equation}
where $\chi$ is a controllable phase. Since there is no spin analyzer before D$_2$, the detection of a neutron in D$_2$ is associated to the projector
\begin{equation}\label{pi2}
	\hat{\Pi}_2=\Big(|I\rangle-\rme^{\rmi\chi}|II\rangle\Big)\Big(\langle I| - \rme^{-\rmi\chi}\langle II|\Big)\otimes \hat{I}_s,
\end{equation}
where $\hat{I}_s$ corresponds to the identity operator for the spin degree of freedom.

In the first part of the experiments, Denkmayr \textit{et al.} place absorbers in each path and see if the detection counts in D$_1$ reduce. Since D$_1$ only detects neutrons with spin state $|-\rangle$ and according to (\ref{psin}) there is no  $|-\rangle$ component for the part of the neutron wavefunction that propagates through arm $I$, the detections should not vary when an absorber is placed in path $I$ and should decrease when the absorber is placed in path $II$. This behavior is observed in the experiments \cite{denkmayr14}.

In the second part of the experiments, magnetic fields are applied in the interferometer paths to produce a small rotation of the neutrons spin. The phase $\chi$ of the projectors from (\ref{pi1}) and (\ref{pi2}) is varied and it is observed if the detection counts in D$_1$ and D$_2$ depend or not on $\chi$. When no magnetic field is applied, there can be no interference since the spin states in each path are orthogonal for the state (\ref{psin}), so the counts on both detectors should not depend on $\chi$. When a magnetic field  is applied in path $I$ changing the neutron spin state from $|+\rangle$ to $a|+\rangle + b|-\rangle$ with $|a|^2+|b|^2=1$, the $|-\rangle$ component of the wavefunction of this path can interfere with the wavefunction of path $II$, such that the counts in both detectors should depend on $\chi$. When a magnetic field  is applied in path $II$ changing the neutron spin state from $|-\rangle$ to $c|-\rangle + d|+\rangle$ with $|c|^2+|d|^2=1$, the $|+\rangle$ component of the wavefunction of this path can interfere with the wavefunction of path $I$, such that the counts in detector D$_2$ should depend on $\chi$. But since the detector D$_1$ selects only the $|-\rangle$ component of spin, the counts in this detector should not depend on $\chi$. All these predictions are confirmed by the experiments \cite{denkmayr14}. 

The behaviors described above led the authors to say: `` (...) an absorber with high transmissivity has on average no significant effect on the measurement outcome if it is placed in path $I$. It is only effective if it is placed in path $II$. In contrast to that, a small magnetic field has on average a significant effect only in path $I$, while it has none in path $II$. Therefore, any probe system that interacts with the Cheshire Cat system weakly enough will on average be affected as if the neutron and its spin are spatially separated''  \cite{denkmayr14}. This is what they mean when they say that  ``the experimental results suggest that the system behaves as if the neutrons go through one beam path, while their magnetic moment travels along the other'' \cite{denkmayr14}. As we have seen here, the results can be explained as simple quantum interference, with no separation between the neutron and its spin. {There is no need of interpreting the results as if the neutron and its spin are spatially separated.}

\section{Final remarks}
\par {In their papers, both Aharonov \textit{et al.} \cite{aharonov13} and Denkmayr \textit{et al.} \cite{denkmayr14} suggest a potential use of the quantum Cheshire Cat. In \cite{aharonov13}, the authors state: ``suppose that we wish to
perform a measurement in which the magnetic moment plays the central role, whilst the charge
causes unwanted disturbances. The question that arises is whether it might be possible to remove
this disturbance, in a post-selected manner, by producing a Cheshire Cat where the charge is
confined to a region of the experiment far from the magnetic moment.'' In the last paragraph of the Discussion in \cite{denkmayr14} there is an analogous proposal. In our framework, though, it is clear that disturbances in any degree of freedom of either the pre- or the post-selection state will change the final state of the pointer, due to a change of the amplitude or phase of each component in (\ref{projection}) -- and therefore (\ref{approx}) will be different. The randomness of these disturbances leads to decoherence of the pointer state, therefore messing up the results.}
\par In our opinion, the paradoxical conclusions that a photon may be separated from its polarization \cite{aharonov13,bancal14} or that a neutron can be separated from its spin \cite{denkmayr14} presented as the `quantum Cheshire Cat' effect are one more apparent paradox that arises whenever we attribute physical reality to quantum superposition states, for instance when describing a quantum particle inside an interferometer prior to its detection \cite{gibbins}. {Recent works have also discussed the quantum Cheshire Cat, specifically the results of Denkmayr \textit{et al.} \cite{stuckey14,atherton15}.} Naive interpretations of delayed choice experiments \cite{wheeler78,jacques07} or quantum erasers \cite{scully91,herzog95,durr98,walborn02} lead to similar apparent paradoxes. As we have shown here, the predictions of \cite{aharonov13} and the experiments of \cite{denkmayr14} can be understood as simple quantum interference, with no separation between the quantum particles and their internal degrees of freedom, and we hope our results provide a better understanding of the phenomenon reinforcing that no interpretation more strange than standard quantum mechanics is required.

\ack

We would like to ackowledge Cristhiano Duarte for useful comments on the manuscript. This work was supported by the Brazilian agencies CNPq, CAPES, FAPEMIG, and PRPq/UFMG.

\section*{References}


\end{document}